# Novel Nanoscroll Structures from Carbon Nitride Layers

E. Perim[*] and D. S. Galvao

* eric.perim@gmail.com

*Applied Physics Department, State University of Campinas, Campinas-SP 13083-959, Brazil.*

## Abstract

Nanoscrolls (papyrus-like nanostructures) are very attractive structures for a variety of applications, due to their tunable diameter values and large accessible surface area. They have been successfully synthesized from different materials. In this work we have investigated, through fully atomistic molecular dynamics simulations, the dynamics of scroll formation for a series of graphene –like carbon nitride (CN) two-dimensional systems: g-CN, triazine-based g-$C_3N_4$, and heptazine-based g-$C_3N_4$. Our results show that stable nanoscrolls can be formed for all of these structures. Possible synthetic routes to produce these nanostructures are also addressed.

# Introduction

Nanoscrolls consist of sheets rolled up in a papyrus-like way, as show in Figure . These are very interesting nanostructures since their open ends provide great radial flexibility and large solvent accessible surface area[1,2], unlike their structurally-related nanotubes. This topological condition makes nanoscroll very promising structures for a large variety of applications, such as, nano-electrical[3] and mechanical actuators[4,5] and hydrogen storage[6,7,8]. Another important aspect of nanoscrolls is that they can, in principle, be made of different layered materials, like carbon[9], boron nitride (BN)[10], niobium oxide[11], potassium niobate[12] and others.

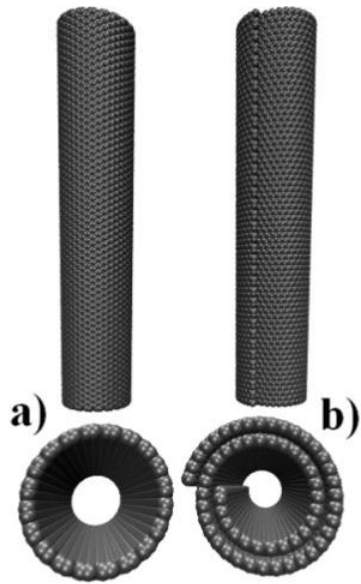

**Figure 1 -** Lateral and top views of: (a) Carbon Nanotubes and; (b) Carbon Nanoscrolls.

Despite carbon whiskers being first reported by Bacon in 1960[13], reliable carbon nanoscrolls (CNNSs) synthesis methods only came out decades later[14,15,16,17], while BN nanoscrolls (BNNSs) were even more recently successfully produced[18,19], a few years after their theoretical prediction[10]. These facts are suggestive that there are still many possible nanoscroll structures waiting to be discovered.

Carbon nitride (CN) structures have been attracting great attention since their prediction as superhard materials[20], leading to many theoretical[21,22,23] and experimental[24,25,26,27] studies. Recently, graphene-like carbon nitride (g-CN) structures have been synthesized[28] with distinct stoichiometries and morphologies (see Figure 2). CN structures, beyond their superhardness, present other interesting properties, such as low density, biocompatibility and interesting electrical and optical properties[29]. By combining these unique CN characteristics with the structural properties inherent to nanoscrolls new nanostructures with very attractive mechanical and electronic properties could be formed. One of the objectives of the present work is to address some of these aspects. In this work we have investigated, through fully atomistic molecular dynamics (MD) simulations (see Experimental Section for details), the stability and the structural and dynamical properties of hypothetical carbon nitride nanoscrolls (CNNSs) formed from rolling up graphene-like layers of the structures shown in Figure 2:

- Carbon nitride layers (g-CN).
- Triazine-based g-$C_3N_4$ (tr-$C_3N_4$) layers (henceforward called triazine).
- Heptazine-based g-$C_3N_4$ (hp-$C_3N_4$) layers (henceforward called heptazine).

Structurally, nanoscrolls are defined by their parent sheet dimensions (L and W) and scrolling angle ($\theta$) (see Figure 3). They can be formed in two different types: alfa-type, when the sheet is scrolled from both ends, and; beta-type, when the sheet is scrolled from one end, the other remaining planar, as schematically shown in Figure 3. It has been shown[9,10] that for a critical rolling diameter the scrolled structure becomes stable. The general process of producing nanoscrolls consists of rolling up planar (or quasi-planar) structures into papyrus-like configurations (Figure 3). The competing forces involved are the elastic deformations and the van der Waals. Depending on some parameters, the scroll can even be more stable than their planar parent structures. Also, beyond a critical point the scrolling process is self-sustained and the resulting scrolled structures oscillate radially near their maximum stability diameter.

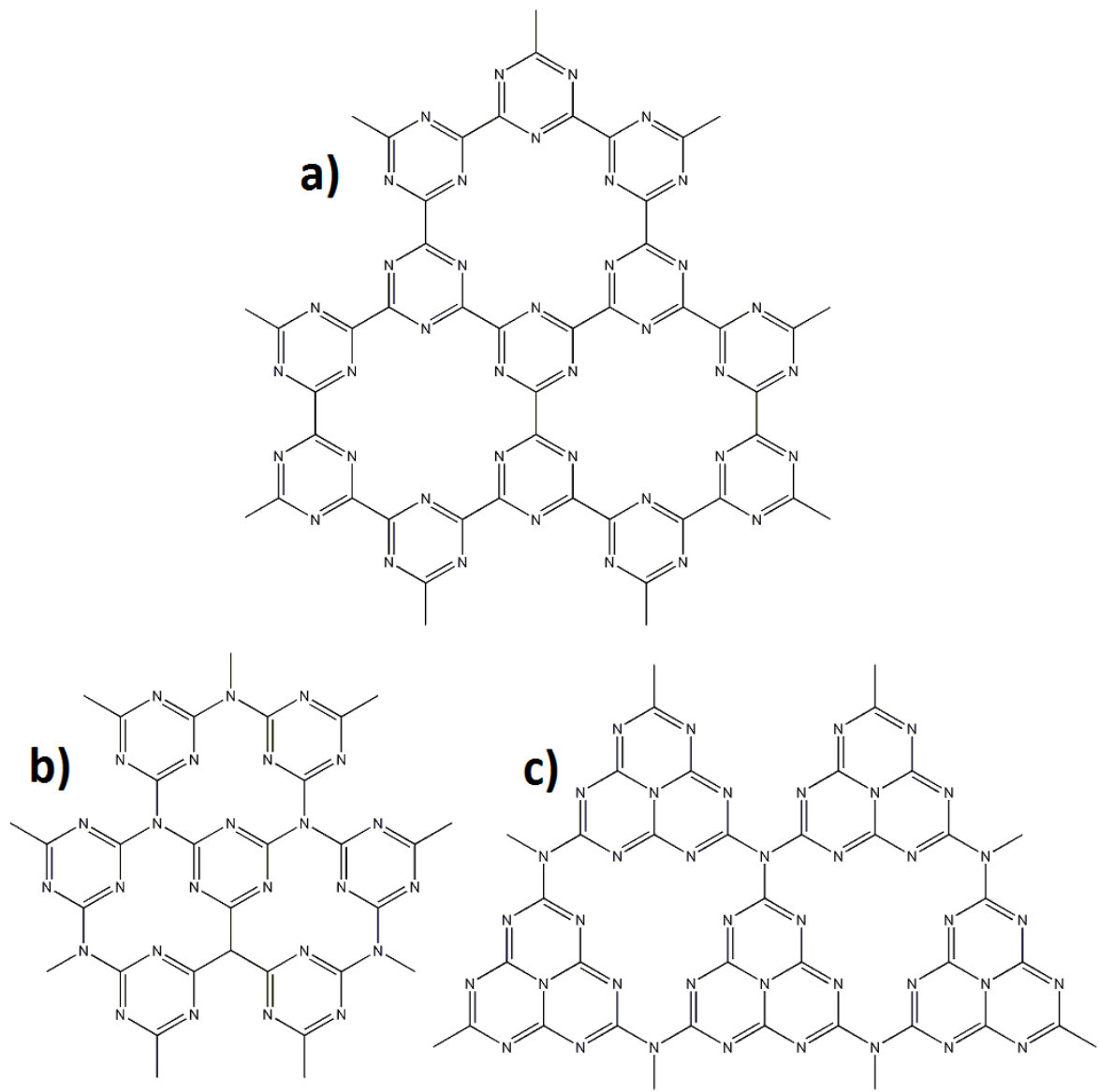

**Figure 2** - Different graphene-like carbon nitride layer structures: a) g-CN ; b) triazine-based g-$C_3N_4$ ; c) heptazine-based g-$C_3N_4$.

With relation to their synthesis, many different experimental techniques have been used[14,15,16] and, recently, some theoretical works have proposed new possible synthetic routes to produce nanoscrolls, such as, inducing the scrolling process to start from direct contact with nanotubes[30,31,32].

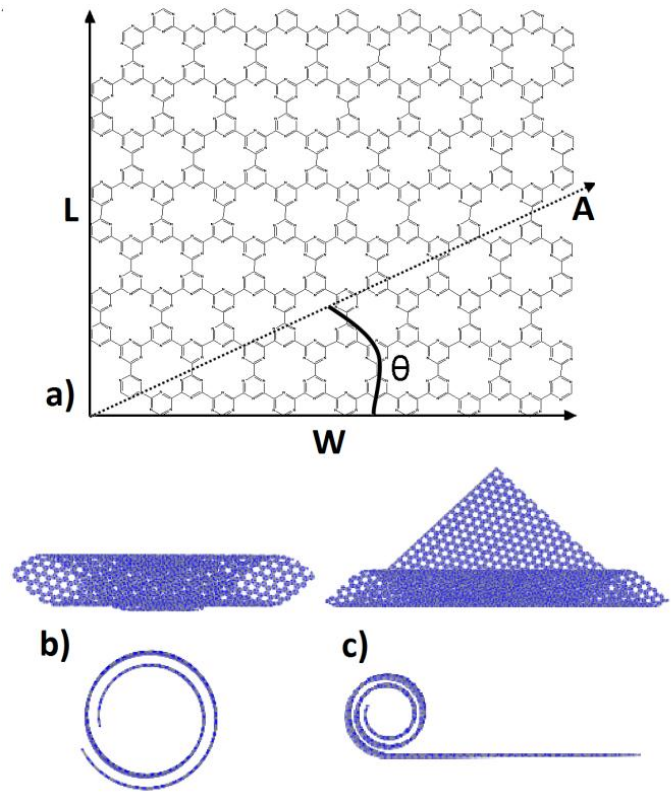

**Figure 3 -** a) Unwrapped g-CN sheet of length L and width W. The θ angle defines the scrolling axis A. b) Alpha-type nanoscroll wtih θ=45°. c) Beta-type nanoscroll with θ=45°.

## Results and Discussion

CNNSs can become stable structures through interplay between van der Waals and elastic forces, as previously observed on carbon[9] and boron nitride[10] scrolls. As a plane sheet is bent from its initial state, energy system increased due to elastic deformations. This process needs to be energy assisted (sonication, for instance), or the sheet would return to its planar (or quasi-planar) conformation. However, if the bending continues up to a point where the sheet ends start to overlap, the van der Waals forces oppose the elastic ones, increasing the structural stability by decreasing the total system energy. Depending on the elasticity of the membrane and the inner diameter value, after the starting of the overlap process, the scrolling process can be self-sustained and a stable scroll can be formed.

In order to investigate the stability and dynamics of formation of these structures, we define a quantity $\Delta E$ as the total energy minus the energy of the planar configuration, divided by the number of atoms. The relation between $\Delta E$ and the scroll diameter is presented in Figure 4.

In Figure 4(a) we present $\Delta E$ as a function of internal scroll diameter values, for CNSs and g-CN nanoscrolls with the same W and L dimensions (as defined in Figure 3). While the global behavior is quite similar, the stability depth well is quite deeper in the case of CNSs. This can be understood by the fact that, since g-CN sheets are largely porous, unlike graphene, the smaller contact area decreases the interlayer interactions.

In Figure 4(b) $\Delta E$ as a function of the internal scroll diameter values is displayed separately for its van der Waals terms and for the bonded terms. The corresponding configurations for some energy values are also displayed in the Figure.

It can be seen from this Figure that as soon as overlap starts to occur, the van der Waals energy sharply decreases, thus compensating the increase in the elastic energy up to the point at which the inner diameter becomes too small and elastic forces once again dominate the global behavior. This region at which the van der Waals forces are dominating defines the stability valley for a nanoscroll. In Figures 4(c) and 4(d) we present the corresponding results for the triazine and heptazine nanoscrolls, respectively. Their global behaviors are quite similar and can be explained by the same arguments. The larger W dimension can explain the deeper stability valley present in Figure 4(d).

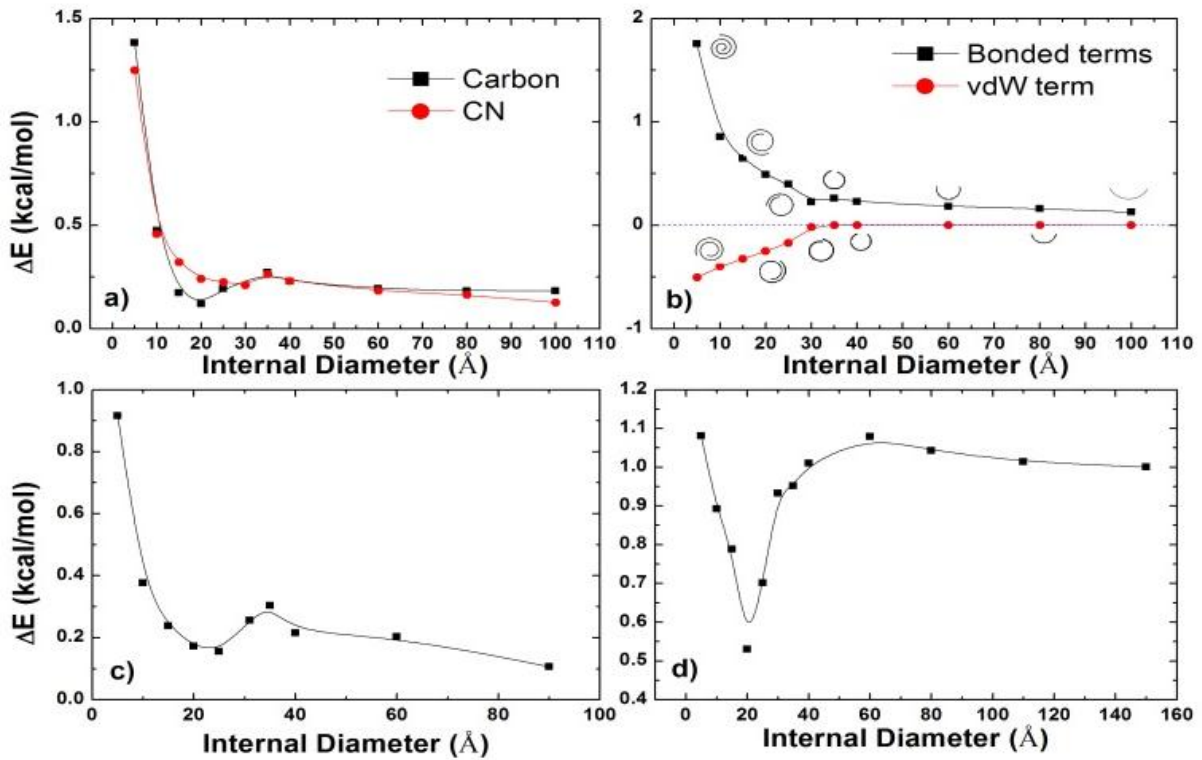

**Figure 4** - ΔE as a function of the internal diameter values for the different scrolls. a) Comparison between carbon and g-CN nanoscrolls, for structures with W~L~100Å. b) ΔE for van der Waals and bonded terms for g-CN nanoscrolls, with a scroll of

W~L~100Å. c) Results for triazine nanoscrolls with W~L~105Å. d) Results for heptazine nanoscrolls with W~ 170Å and L~107Å. All cases for θ=0º.

For a fixed scrolling angle of θ=0º, the W dimension defines the number of layers of the scrolled structure, while the L dimension defines its length. The influence of scroll dimensions on its stability can be better understood analyzing the curves in Figures 5 and 6. By fixing one dimension and varying the other with a fixed internal diameter (of ~20Å), we can isolate the contribution of each scroll dimension. In Figure 5 we can see that larger W means greater stability, i.e., a deeper stability valley, as seen on Figure 4(d). As W increases, more layers are added to the scroll, thus resulting in stronger interlayer interactions and, consequently, higher stability. On the other hand, the L dimension values have almost no effect on the energy stability, except for extremely small values, as shown in **Error! Reference source not found.**. This means that scroll length does not significantly affect their stability.

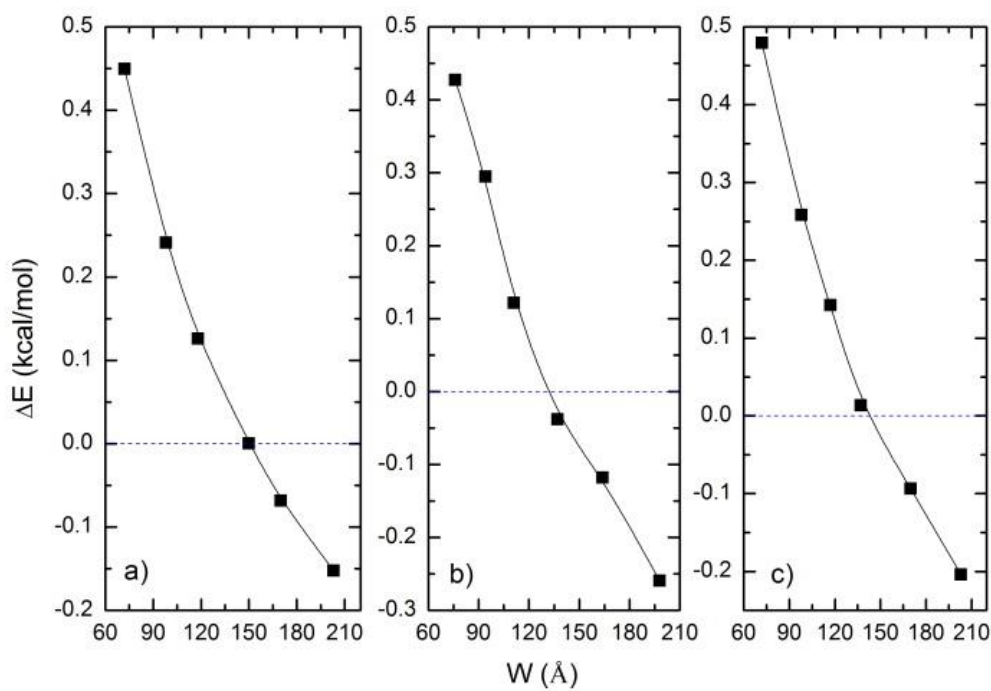

**Figure 5** - ΔE as a function of W values. Results for: (a) g-CN; (b) triazine, and; (c) heptazine nanoscrolls. For all cases θ=0º.

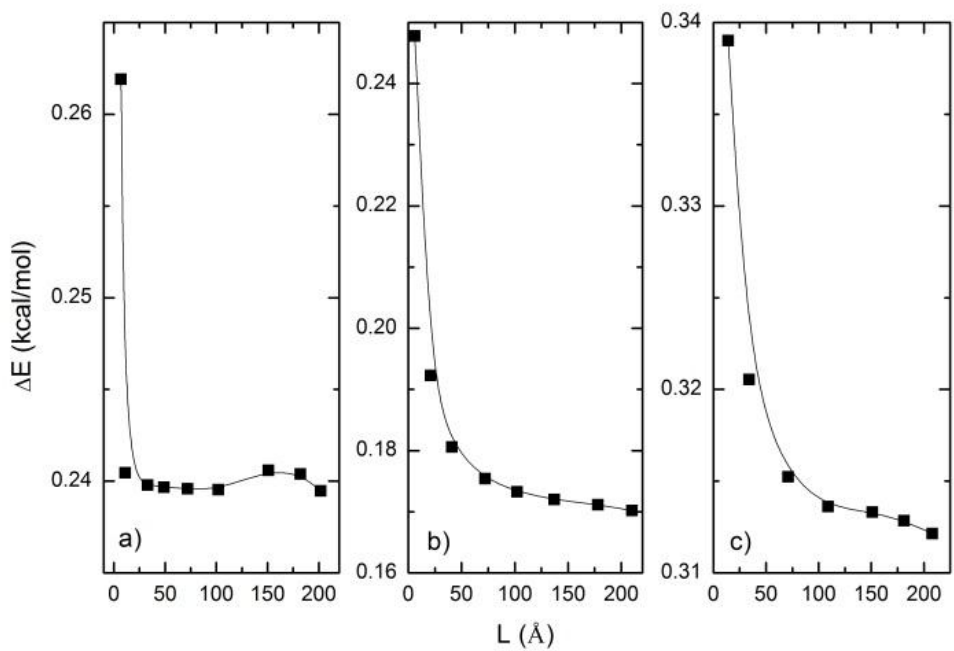

**Figure 6** - ΔE as a function of L values. Results for: (a) g-CN; (b) triazine, and; (c) heptazine nanoscrolls. For all cases θ=0º.

Different scrolling angles θ can lead to distinct structures in an analogy to nanotubes chiral angles. In order to understand the role of this angle in the structure dynamical properties, we carried out molecular dynamics simulations for alfa-type g-CN nanoscrolls rolled up with three different scrolling angles, θ=0º, θ=90º and θ=45º and inner diameter value near the stability one. ΔE as a function of time evolution for each case is shown in Figure (see Supplementary Information for videos). A very similar behavior between θ=0º and θ=90º structures is noticeable, and they ultimately oscillate around similar ΔE values, indicating no significant influence of the initial θ angle on the final stability, differently from what was reported for BNNSs[10]. For θ=45º, ΔE takes longer to decrease, as the scroll has to undergo significant structural changes to maximize contact area, ultimately converging to either a θ=0º or a θ=90º scroll (see Supplementary Information). For comparison purposes in Figure 7 it is also presented the ΔE time evolution for a beta-type θ=0º g-CN scroll (see Supplementary Information for the dynamics video). Since the beta-type scroll starts with a planar region, it is necessary first to dissipate this elastic tension before the scrolling process can start. This is illustrated by the fact that the ΔE for bonded terms rapidly decreases, while the ΔE for van der Waals terms is still significantly greater, since it takes longer for significant layer overlapping to occur. Thus, it shows that rolling up a sheet from both ends simultaneously (alfa-type) is easier then rolling from one end while the other is kept plane (beta-type). The armchair, zigzag and chiral labels refer to edge-type sheet terminations (see Figure 3) that originate the scrolls.

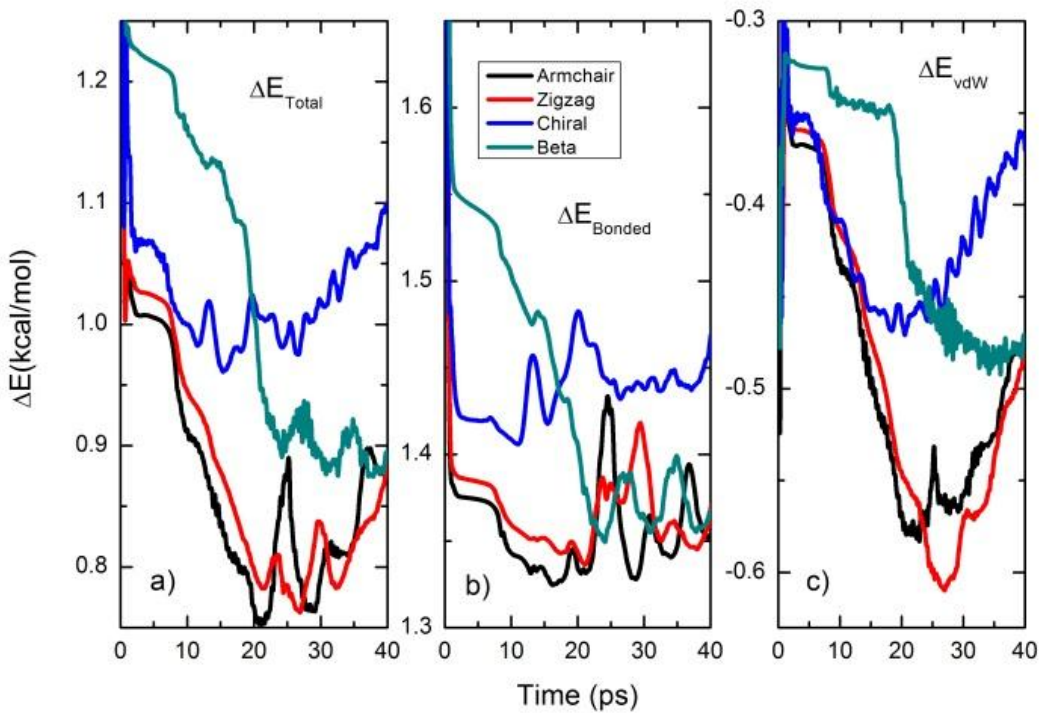

**Figure 7 -** ΔE as a function of the time of simulation: (a) total energy; (b) ΔE bonded values, and; (c) ΔE van der Waals values. Results for alfa-type armchair, zigzag and chiral (with θ=45º) and beta-type armchair g-CN nanoscrolls, respectively.

Lastly, inspired by recently proposed synthesis methods for carbon and boron nitride nanoscrolls in which CNTs are used to initiate sheet scrolling[30,31,32], we have also analyzed the dynamics of g-CN, triazine and heptazine sheets in contact with CNTs. As expected, the tubes induced the start of the scrolling processes of the sheets and they spontaneously formed nanoscrolls. In Figure 8 we show representative screenshots of this process for a g-CN sheet. In the Supplementary Material there are videos showing the dynamics for all three cases. In Figure 9 the ΔE time evolution is displayed for the g-CN case. A sharp decrease in all terms is observed as the structure evolves. These results show that the use of a CNT as an inducer for scrolling is quite effective. These results indicate that this process can be a possible route for the production of these novel nanoscrolls, as carbon nitride planar sheets have already been successfully produced[28].

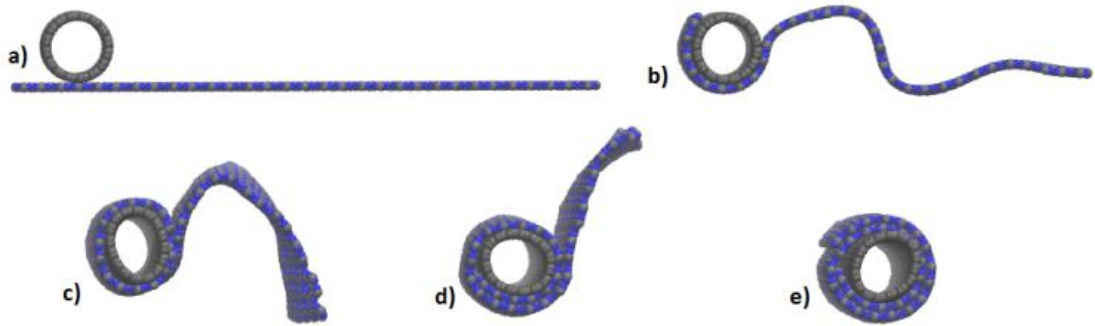

**Figure 8** – Representative snapshots of g-CN nanoscroll formation initiated by contact with a CNT.

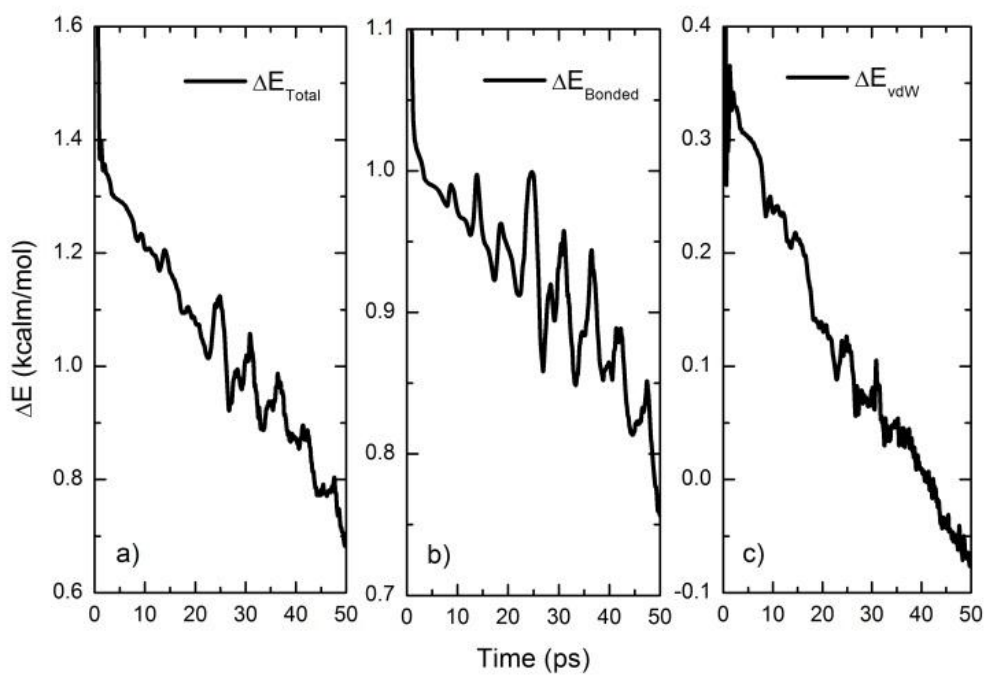

**Figure 9 -** Time evolution of: (a) total ΔE; (b) ΔE bonded terms, and; (c) ΔE van der Waals terms. Results for a g-CN sheet scrolling up induced by a CNT.

## Conclusion

In this work we have investigated, through fully atomistic molecular dynamics simulations, the stability and formation dynamics of carbon nitride (CN) scrolled two-dimensional layers. We have investigated graphene-like CN, triazine-based $g-C_3N_4$ and heptazine-based $g-C_3N_4$ structures. Our results show that stable nanoscrolls can be formed for all of these structures. We have also demonstrated that carbon nanotubes can be effectively used to induce the start of the scrolling processes. As the CN sheets have been already synthesized, the production of scrolled structures is perfectly feasible with our present-day technology. We hope the present work can stimulate further studies along these lines.

## Experimental Section

All calculations were carried out using the Universal Force Field[33] as implemented on the Accelrys Materials Studio Suite[34], which has already been used to successfully study nanoscrolls[9,10]. All atoms were assumed to be neutral with partial double bonds and no charge optimization was carried. Some tests carried with ReaxFF lead to unphysical results (some bent, non‐overlapping, sheets being more stable than their plane counterparts). Similar results were observed for UFF with charge equilibration. Dynamics were simulated with a time step of 1fs and a NVT ensemble with temperature controlled at 50K by a Nosé-Hover thermostat. Scrolls were generated by rolling up two dimensional sheets into truncated Archimedean-type spirals.

Considering that the C-N bond could be highly polarizable[35,36] and we are using no charge equilibration in our calculations, we have also carried out benchmark calculations on beta-$C_3N_4$ crystals and the results were contrasted again the ab initio ones reported in ref. 37. We compared the lattice parameters a and c, and three different C-N bond lengths. The ab initio/classical values, in angstroms, were respectively 5.41/5.51, 2.40/2.49, 1.46/1.47, 1.46/1.49 and 1.45/1.48, the deviation being only of 2%, 4%, 1%, 2% and 2% respectively, thus indicating the adequacy of this approach to describing carbon nitride structures.

## Acknowledgements

*This work was supported in part by the Brazilian Agencies CNPq, CAPES and FAPESP. The authors thank the Center for Computational Engineering and Sciences at Unicamp for financial support through the FAPESP/CEPID Grant #2013/08293-7.*